\documentclass[%
 reprint,
 superscriptaddress,
 longbibliography,
 amsmath,amssymb,
 aps,
 pra,
]{revtex4-2}

\usepackage{graphicx}
\usepackage{dcolumn}
\usepackage{bm}
\usepackage{mathtools}
\usepackage{hyperref}

\begin{document}

\preprint{APS/123-QED}

\title{Integrated squeezed light sources for two-mode entanglement \\ in thin-film lithium niobate}

\newcommand{\bigq}{
Center for Macroscopic Quantum States bigQ, Department of Physics, Technical University of Denmark, Fysikvej 307, DK-2800 Kgs. Lyngby, Denmark
}

\newcommand{\uhei}{
Kirchhoff-Institute for Physics, Heidelberg University, Heidelberg, Germany
}

\author{Philipp Lohmann}
\thanks{These authors contributed equally to this work. \\ 
Correspondence should be addressed to Renato R. Domeneguetti. Email: rerdo@dtu.dk}
\affiliation{\uhei}

\author{Renato R. Domeneguetti}
\thanks{These authors contributed equally to this work. \\
Correspondence should be addressed to Renato R. Domeneguetti. Email: rerdo@dtu.dk}
\affiliation{\bigq}

\author{Daniel Wendland}
\affiliation{\uhei}
\author{Alessandro Perino}
\affiliation{\bigq}
\author{Tobias Egebjerg}
\affiliation{\bigq}
\author{Liam McRae}
\affiliation{\uhei}
\author{Jonas S. Neergaard-Nielsen}
\affiliation{\bigq}
\author{Wolfram H.P. Pernice}
\affiliation{\uhei}
\author{Ulrik L. Andersen}
\affiliation{\bigq}

\date{\today}

\begin{abstract}
Scalable generation of nonclassical light sources on an integrated platform is a key requirement for photonic quantum information processing. In particular, realizing multiple indistinguishable squeezed light sources on a single chip is an essential step toward continuous-variable quantum computing. Here, we demonstrate the fabrication of two indistinguishable and independently controllable optical parametric oscillators on a thin-film lithium niobate (TFLN) platform. The device design focuses on reproducibility, independent tunability, and compatibility with larger telecom-wavelength continuous-variable photonic circuits. We observe up to 0.5 dB of directly measured squeezing below the shot-noise level from each source. By interfering the two modes on a beam splitter, we generate an EPR-type two-mode squeezed state and verify continuous-variable entanglement through violation of the Duan–Simon inseparability criterion. This is the first demonstration of two independently tunable squeezed-light sources on a single TFLN chip and their use for generating continuous-variable entanglement.
\end{abstract}


\maketitle


\section{Introduction}

Squeezed states of light constitute a fundamental resource in quantum optics, enabling the reduction of quantum noise in one field quadrature below the standard quantum limit at the expense of increased noise in the conjugate quadrature. Such nonclassical states have found widespread applications in quantum information processing, quantum communication, and precision metrology, including enhanced interferometric sensitivity beyond the shot-noise limit~\cite{Andersen2016}.

In particular, continuous-variable (CV) quantum information protocols rely on the deterministic generation of squeezed states, which constitute the key resource for CV measurement-based quantum computing (MBQC), where computation is implemented via adaptive homodyne measurements on large entangled cluster states~\cite{Menicucci2011_TemporalModeCVClusters}. One main advantage of this approach is that a finite, fixed amount of resources can generate cluster states of indefinite size by encoding quantum information in time on demand.  Although one- and two-dimensional CV cluster states can be realized in bulk optics~\cite{Yokoyama2013_TimeDomainCVCluster, Asavanant2019_2D_TDM_CVCluster, Larsen2019_2D_Deterministic}, cluster states as required for large computation are limited by the coherence time of the laser, losses, and phase lock stability.

Integrated photonics is a natural route to realize a more efficient and robust hardware for fault-tolerant quantum computing and quantum sensing. Photonic approaches offer highly efficient sources of squeezed states, intrinsic phase stability, low propagation losses and efficient in- and outcoupling methods, which renders integrated photonics a promising platform especially for CV-based quantum computation~\cite{AghaeeRad2025_ScalingModularPhotonicQC,Madsen2022}.

The currently available platforms can be distinguished by the nonlinear processes they support. Squeezed light can be produced by four-wave-mixing (FWM), which exploits the third-order nonlinearity, whose weak effective nonlinear interaction is enhanced by using ring resonators with ultra-high quality factors. Passive phase matching occurs mainly through dispersion engineering of the waveguide geometry and Kerr-type interactions~\cite{levy2010cmos,Gong:20}. 
Silicon nitride as a mature and low-loss platform has been established as the main platform for squeezed light generation via FWM inside a microresonator~\cite{dutt2014,Shen:25,Ulanov2025,tritschler2025chipintegratedsinglemodecoherentsqueezedlight,larsen2025gkp}. However, the creation of entanglement is difficult due to excess noise~\cite{AlfredoKogler:24}, thermal noise~\cite{thermo-noise} and other nonlinear processes like self- and cross-phase modulation, which are difficult to mitigate without increasing the complexity of the optical setup~\cite{zhang2021}.

\begin{figure*}[!ht]
\centering
\includegraphics[width=\textwidth]{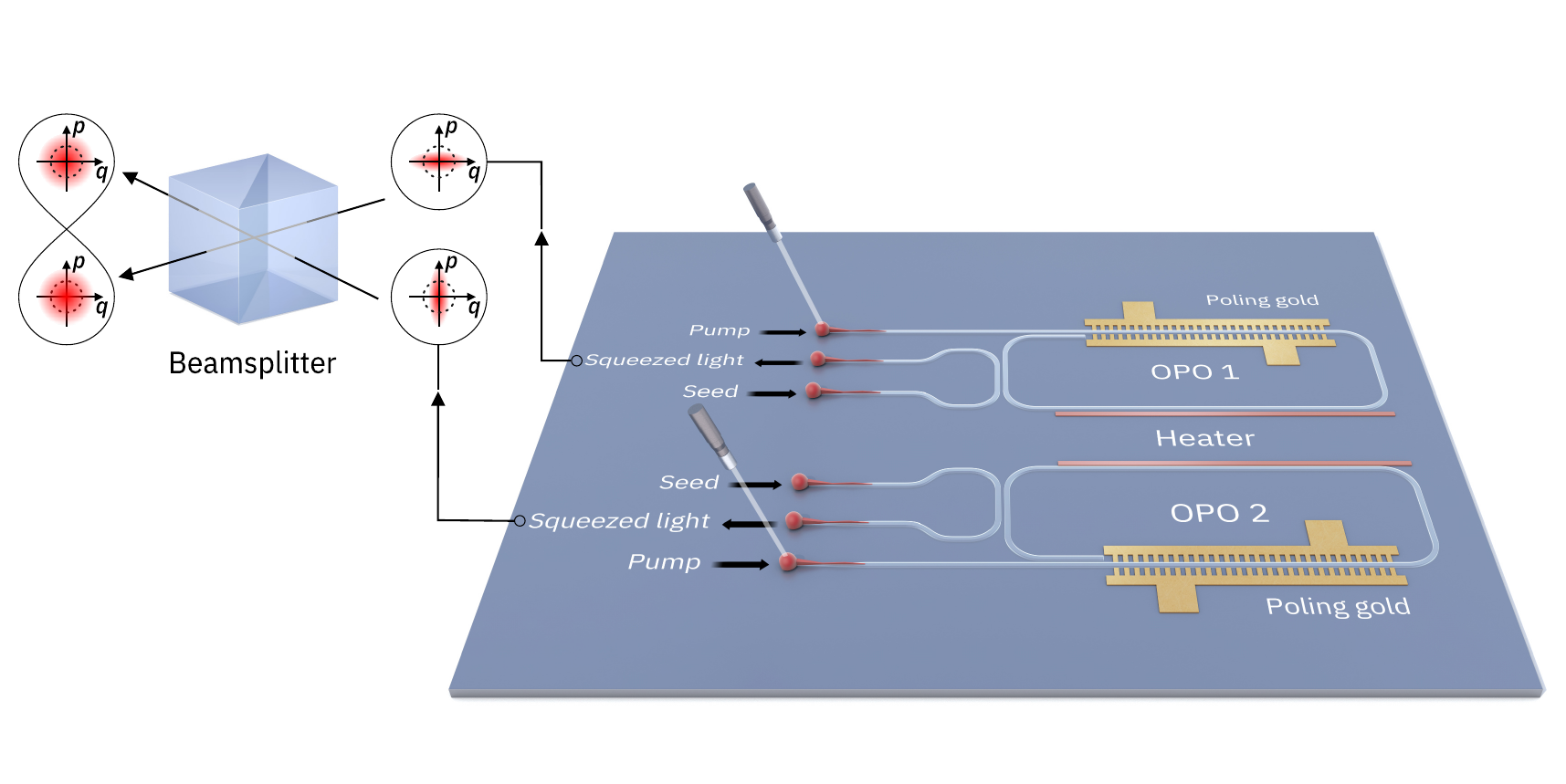}
\caption{Scheme of the integrated two-source squeezed-light device and off-chip two-mode entanglement generation. Two single-resonant OPOs are fabricated on a TFLN chip. Each OPO contains a periodically poled section, an on-chip heater for resonance tuning, and a dichroic directional coupler for separating pump and signal light. The squeezed outputs are collected from the chip and interfered on an off-chip beam splitter to generate a two-mode squeezed state. Seed inputs are used to lock the squeezing quadratures. The dashed circle represents the shot noise variance reference level.
}
\label{fig:fig_1}
\end{figure*}

For materials with a strong second-order nonlinearity, squeezed light can be generated via parametric down-conversion (PDC), offering higher efficiency, lower noise, and greater flexibility. Due to its strong nonlinearity, mature fabrication for integrated circuits and a large transparency window, lithium niobate is one of the most promising candidates for complex quantum circuitry \cite{Weis1985}. In addition to PDC processes enabled by periodic poling, thermo- and electro-optic modulators can be fabricated for linear optical operations~\cite{Zhu:21}. Recently, single-photon detectors have been demonstrated on lithium niobate~\cite{Lomonte2021}, which completes the set of required building blocks for photonic quantum computing.
In waveguides created by reverse proton exchange or titanium diffusion, squeezed light sources with up to 3.1~dB of squeezing have been demonstrated~\cite{Stefszky2017,Mondain:19,domeneguetti2023guided}. The fabrication of a two-mode entangled state generator containing two squeezed light sources with up to 1.38~dB of squeezing has been accomplished, which demonstrates the potential of the platform  \cite{doi:10.1126/sciadv.aat9331}.
However, large building blocks, finite nonlinear efficiency and large bending radii limit the  minimal required footprint and thus the circuit depth, which renders the resulting devices unfeasible for more complex circuits as would be required for quantum applications.

Thin-film lithium niobate (TFLN) with sub-µm$^2$ light confinement can overcome those challenges. 
In recent years, first squeezed light sources based on TFLN were fabricated. In single-pass waveguides, ultrabroadband squeezing \cite{Chen:22,https://doi.org/10.1515/nanoph-2025-0377} and quantum optical phase sensors \cite{Stokowski2023} were realized, demonstrating squeezing ranging from 0.12 to 1.4 dB. By harnessing the strong peak power of a 75 fs pulsed laser source, squeezing close to 4 dB at a wavelength around 2~µm was obtained with an optical parametric amplifier~\cite{doi:10.1126/science.abo6213}. 
Exploiting the local enhancement of a double-resonant optical-parametric oscillator (OPO), 0.46~dB of squeezing was measured in Z-cut TFLN using modal phase matching ~\cite{Arge2025_TFLN_squeezing}, whereas 0.81~dB of squeezing was measured for quasi-phase-matching in X-cut TFLN~\cite{ren2026quantumsqueezingallresonantperiodically}.

By using a second harmonic generation (SHG) process combined with a single-resonant OPO, the generation of pump light, a squeezed light source of 0.55~dB squeezing and mixing with an on-chip local oscillator were shown, which demonstrates the co-integration of key components for CV quantum computing on a single chip \cite{doi:10.1126/sciadv.adl1814}. 

All of the above mentioned works on TFLN focus on the realization of a single squeezed light source. The demonstration of a robust approach towards multiple sources, which is required by most CV quantum computing schemes, is still missing. Reproducible nonlinear processes at a predefined phase-matching wavelength along with individual tunability of all components is mandatory.

This work demonstrates the first realization of two integrated and indistinguishable squeezed light sources based on tunable single-resonant OPOs on a single TFLN chip. Squeezing close to 0.5 dB is measured around the target wavelength of 1550 nm in both sources. 
As a first step toward integrated CV cluster-state generation, we validate the mutual compatibility, spectral matching, and phase coherence of the two sources by interfering them to generate a two-mode squeezed state. Continuous-variable entanglement is verified through violation of the Duan–Simon inseparability criterion.

\section{Methods}
\subsection{Design}

We design the two-mode squeezed light generator closely connected to the fabrication limitations of the TFLN platform. A tradeoff between limited pump power, optical losses, fabrication robustness and limited nonlinear conversion efficiency has to be found. Additionally, the operating wavelength is predetermined, which guarantees future compatibility with larger integrated quantum photonic circuits. Motivated by the tunability of high precision laser systems, low losses and technical maturity in the near infrared c-band, the desired degenerate SPDC  wavelength is set to 1550 $\pm$ 0.5 nm, with a corresponding pump at 775 $\pm$ 0.25 nm wavelength. 

A scheme of the chip layout and measurement scheme is demonstrated in Fig~\ref{fig:fig_1}. Two identical squeezed light sources are fabricated using two single-resonant OPOs. Both contain a heater for resonance tuning and a periodically poled region for the nonlinear process. The single-resonance condition is required to ensure operation at the target wavelength. The OPOs are pumped at 775~nm wavelength light through the pump couplers and the resulting quantum state is extracted from the squeezed light ports. By using the respective seed ports, the phase of the squeezing can be locked to the different quadratures. An off-chip beamsplitter is attached to mix both ports and the entangled states can be measured at the corresponding outputs.

\begin{figure*}[!ht]
\centering
\includegraphics[width=\textwidth]{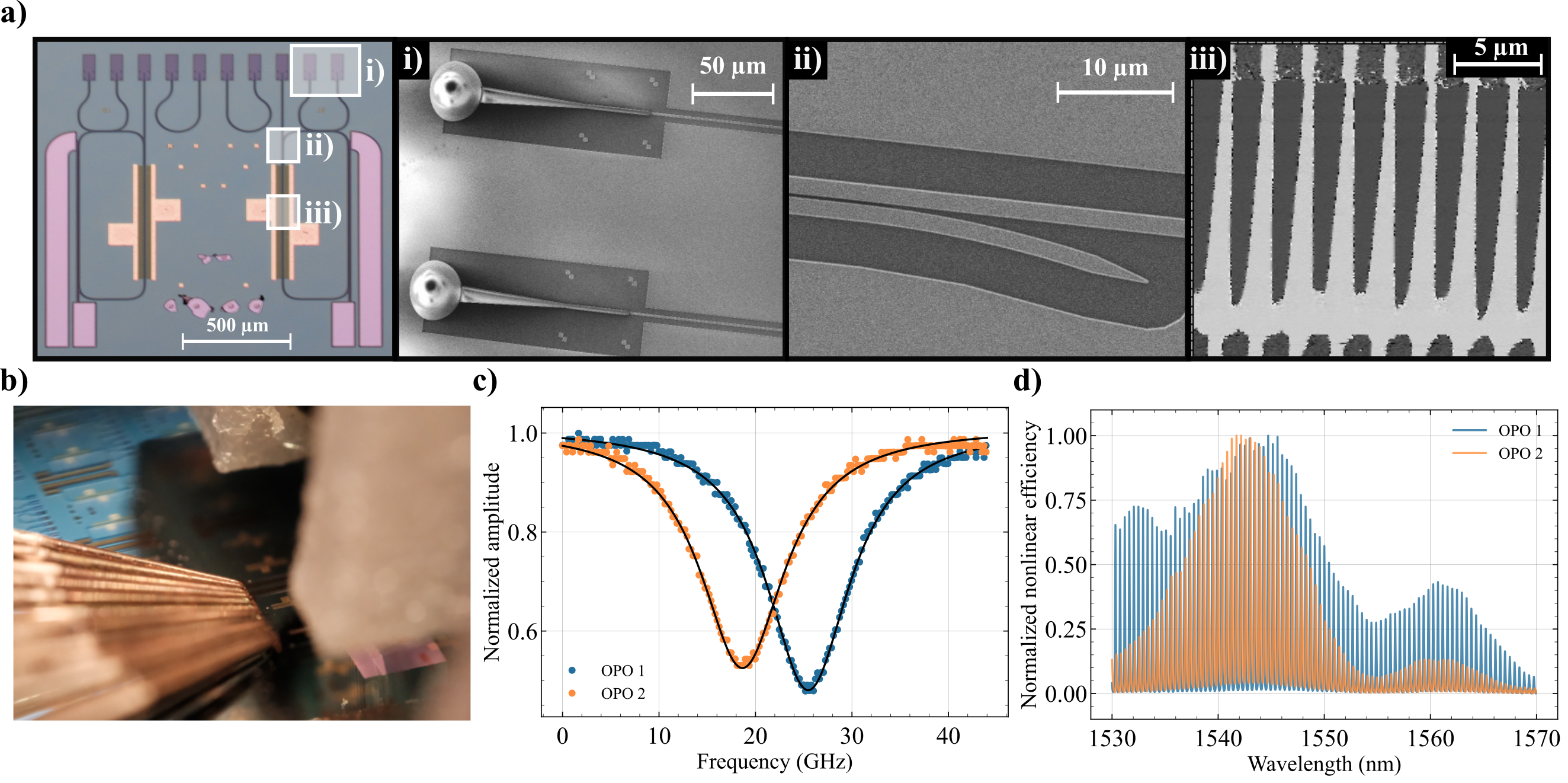}
\caption{ \textbf{a)}  Optical microscope image during the fabrication taken before printing the 3D-couplers on the chip. Details of the integrated platform: \textbf{i)}  SEM image of the fiber-to-chip couplers, \textbf{ii)} SEM image of the dichroic directional coupler used to separate pump and signal light, and \textbf{iii)} piezoresponse force microscopy (PFM) image of the periodically poled lithium niobate before photonic fabrication, confirming uniform domain inversion. 
The waveguide is subsequently etched in the region where the periodic poling has approximately 50~$\%$ duty cycle.
\textbf{b)} Camera image of the fabricated chip showing the fiber array interface and electrical probe contacts.
 \textbf{c)} Measured transmission spectra of the two resonators around 1550 nm wavelength. Loaded quality factors of  \(Q_L = 14.403(9) \times 10^3\) and  \(Q_L = 13.773(10) \times 10^3\) are extracted for OPO 1 and 2 respectively. \textbf{d)} Second-harmonic generation (SHG) spectra used to determine the phase-matching condition. Both resonators have the maximal conversion efficiency around 1543 to 1545 nm wavelength. OPO~2 shows a spectrum close to the optimal sinc-squared shape expected from theory whereas the spectrum of OPO~1 is broadened. The broadened SHG spectrum of OPO~1 is attributed to fabrication-induced variations of the phase-matching condition along the poled section.
}
\label{fig:classical results}
\end{figure*}

An optical microscope image of the chip during fabrication is shown in Fig.~\ref{fig:classical results} \textbf{a)}. The waveguide geometry is based on a fully etched, 300~nm thick TFLN film on 4.7~µm silicon dioxide. The waveguide width is chosen to be 1 µm, which leads to single-mode propagation for the squeezed light and is compatible with low-loss polymer couplers \cite{gehring2019}. A scanning electron microscope (SEM) image of the couplers are shown in subpanel \textbf{i)}.
\noindent Since the waveguide geometry supports many guided modes for the pump light at 775 nm wavelength, a straight geometry is chosen for the path between the coupler and the OPO to reduce additional losses from mode transitions in bent sections of the waveguide. At the entrance to the OPO, the pump light passes a dichroic directional coupler. Due to the difference in evanescence field between signal and pump, the directional coupler can be engineered to cross-couple all the signal light without coupling the pump. As a consequence, the pump light is scattered out of the chip after one round trip in the ring. This design allows for controlled and reproducible cavity resonance tuning, which is crucial for scalable approaches.
A SEM image of the dichroic directional coupler is shown in subpanel~\textbf{ii)}. The dichroic splitter inside the resonator is designed according to prior measurements on a test sample. A gap of 420~nm and a length of 80~µm  is chosen.

On-chip heaters are used to manipulate the refractive index and thus the cavity resonance. The heaters are positioned at opposite sides of the sample to minimize thermal crosstalk. In Fig.~\ref{fig:classical results} \textbf{a)}, the contact pads for the heaters are identified in pink, indicating the electrical access point for the probes at the bottom of the image.

The length of the periodically poled section on the other side of the OPO is chosen as 650 µm, which results in robust and reproducible fabrication. 
Height fluctuations on the nanoscale can change the phase-matching conditions, which has a strong impact on the local resonance condition of the highly dispersive waveguide \cite{Chen2024}. However, the impact is highly reduced on a sub-millimeter long periodically-poled section \cite{Chen2024,Xin2025}. In addition, the limited length results in a wide bandwidth $> 10$ nm in wavelength of the nonlinear process. This allows to compensate fluctuations of 1~nm in waveguide height and 4 nm in width. 
In addition to the considerations described above, a long enough resonator with \(\mathrm{FSR \approx 55~GHz}\) was chosen to reduce bending losses in the curved regions.

The chosen poling length limits the nonlinear conversion efficiency. Nevertheless, with the OPO approach this can be compensated with enough pump power and a sufficiently low-loss cavity. The resulting squeezing $S_-$ and anti-squeezing $S_+$ as a function of pump power $P_{pump}$ is given by 

\begin{equation}
    S_\pm(P_{pump}) = 1\pm4\eta_{tot}\frac{\sqrt{P_{pump}/P_{th}}}{(1\mp\sqrt{P_{pump}/P_{th}})^2},
\label{eq.sqz}
\end{equation}
with \(\eta_{tot} \coloneqq \eta_{esc}\eta_{det}\eta_{cpl}\) describing the cavity escape efficiency $\eta_{esc}$, the out-coupling efficiency from OPO to the fiber $\eta_{cpl}$ and the detection efficiency \(\eta_{det}\).
\noindent The cavity escape efficiency, \(\eta_{esc} = T/(T+L)\), with the outcoupling power fraction per round trip \(T\) and spurious intracavity loss per round-trip \(L\) is controlled by an additional waveguide coupled to the resonator, which is used to extract the squeezed light and additionally serves as an input for a weak seed signal for phase locking. 

\subsection{Fabrication}
Waveguides are defined on 300 nm-thick X-cut TFLN films bonded to a silicon dioxide oxide layer on silicon. The wafer is commercially acquired from NanoLN.

Electrodes for electric field poling (EFP) are patterned using electron beam lithography (EBL) at 100 kV (Raith EBPG5150) with a positive resist. This is followed by physical vapor deposition of an 80~nm-thick gold layer atop a 5~nm-thick chromium adhesion layer, using a lift-off technique. Prior to photonic circuit fabrication, periodic poling is carried out to achieve homogeneous inversion of the ferroelectric domains throughout the entire film thickness. A 15~µm electrode gap and the application of ten sub-millisecond pulses at 395~V ensure uniform poling along the full device length. The poling quality is monitored during the EFP process using piezoresponse force microscopy (PFM) \cite{cryst11030288}. A PFM image of a 20x20 um section of the film is shown in Fig~\ref{fig:classical results} \textbf{a)} subpanel \textbf{iii)} prior to photonic fabrication. The gold fingers are visible on the top and bottom of the image. Domain inversion starts at the top side, where a positive voltage is applied and the poled domains are visualized by the black regions in the center of the image. The trapezoidal shape guarantees a region with a 50 percent duty cycle which is chosen for further fabrication.

Waveguides are subsequently patterned using a positive-tone resist (AR-P 6200.13), EBL, and etching via argon milling (Oxford Instruments PlasmaPro 100 ICP). An RCA-1 clean removes redeposited material from the waveguide sidewalls, and an annealing step minimizes crystal defects.

The heater electrodes are patterned using the same approach as for the poling gold. Tungsten is chosen as a heater metal and deposited using a sputter machine. Finally, fiber-to-chip couplers are fabricated using  a Nanoscribe QX system and direct laser lithography. 

\section{Classical characterization}

Prior to squeezing measurements, the fabricated on-chip OPOs are characterized and their suitability for quantum measurements is analyzed. By using a Santec TSL-770 tunable light source, the infrared transmission spectrum gives insight into the ring resonances and losses whereas the SHG spectrum is measured to obtain the resonance of the nonlinear process.

In Fig.~\ref{fig:classical results} \textbf{b)} a camera image of the chip within the measurement setup is shown. The optical interface is realized with a single fiber array (pink reflection on the right side) controlling all optical in- and outputs, whereas the electrical signals are connected to the chip using a single electrical probe.

The resulting ring resonances around a central wavelength of 1550~nm of both OPOs are shown in Fig.~\ref{fig:classical results}~\textbf{c)}. Slight deviations in resonance wavelength, extinction ratio, and bandwidth are obtained between both resonators, which can be attributed to fluctuations in the fabrication process. The two resonance frequencies differ by nearly $10$~GHz, which is compensated in further measurements by the on-chip heaters. The half width at half maximum (HWHM) bandwidths of resonators 1 and 2 are measured to be \(6.719 \pm 0.004\) and \(7.026 \pm 0.005~\mathrm{GHz}\), and consequently, the loaded quality factors are \(14.403 \pm 0.009\) and \(13.773 \pm 0.010 \times 10^3\), respectively. These parameters indicate strongly overcoupled resonators and the escape efficiencies can be obtained from the extinction ratios as \(\eta_{esc}^{(1)} = 0.846 \pm 0.004\) and  \(\eta_{esc}^{(2)} = 0.864 \pm 0.003\). Additionally, the fiber-to-chip coupling losses are measured as approximately \(1.5~\mathrm{dB}\) at \(1550~\mathrm{nm}\) and \(4~\mathrm{dB}\) at \(775~\mathrm{nm}\) yielding for both OPOs \(\eta_{cpl}^{(1)} = \eta_{cpl}^{(2)} = 0.72 \pm 0.07\). The detection efficiencies are determined as \(\eta_{det}^{(1)} = \eta_{det}^{(2)} = 0.87 \pm 0.07\), which includes \(\mathrm{95~\pm~5}\)\% photodiode quantum efficiency and additional 5\% loss coming from fiber sleeves and \(\mathrm{1\%}\) splitting used for seed locking, as discussed in the following paragraphs. The interference visibility between seed and local oscillator fields was measured to be \(\mathrm{\ge 99~\%}\), confirming high spatial and polarization mode overlap. The corresponding contribution to the detection inefficiency is therefore negligible compared with the coupling and detector losses. The only difference between both paths comes from escape efficiencies, leading to \(\eta_{tot}^{(1)} = 0.53 \pm 0.07\) and \(\eta_{tot}^{(2)} = 0.54 \pm 0.07\). The measured parameters are summarized in Table~\ref{tab:opos}.

The nonlinear resonance is extracted from the SHG transmission of the pump port and the normalized spectrum is shown in Fig.~\ref{fig:classical results}~\textbf{d)}. The maximal peak conversion efficiency is found between \(1543\) and \(1545~\mathrm{nm}\) wavelength at room temperature, which is below the target of \(1550~\mathrm{nm}\). Therefore, moderate heating with a Peltier element placed beneath the chip allows for adjustment of the overall chip temperature, which is used to compensate for the deviation from the desired wavelength of the nonlinear process. A tunability of roughly 0.1 nm/K is obtained. 

\begin{table}[h]
\caption{\label{tab:opos}Measured parameters at 1550~nm for the two OPOs.}
\begin{ruledtabular}
\begin{tabular}{lcc}
Parameter& OPO~1& OPO~2\\
\hline
& & \\[-6pt]
\(\frac{1}{2}\)-bandwidth (GHz) & 6.719(4) & 7.026(5) \\[3pt]
Loaded-\(Q\) (\(\times 10^3\)) & 14.403(9) & 13.773(10) \\[3pt]
Escape efficiency & 0.846(4) & 0.864(3) \\[3pt]
Fiber-to-chip efficiency per facet & 0.72(7) & 0.72(7) \\[3pt]
Detection efficiency & 0.87(7) &  0.87(7) \\[3pt]
Total efficiency & 0.53(7) & 0.54(7) \\
\end{tabular}
\end{ruledtabular}
\end{table}

\section{Quantum measurements}
\begin{figure*}[!ht]
\centering
\includegraphics[width=0.9\textwidth]{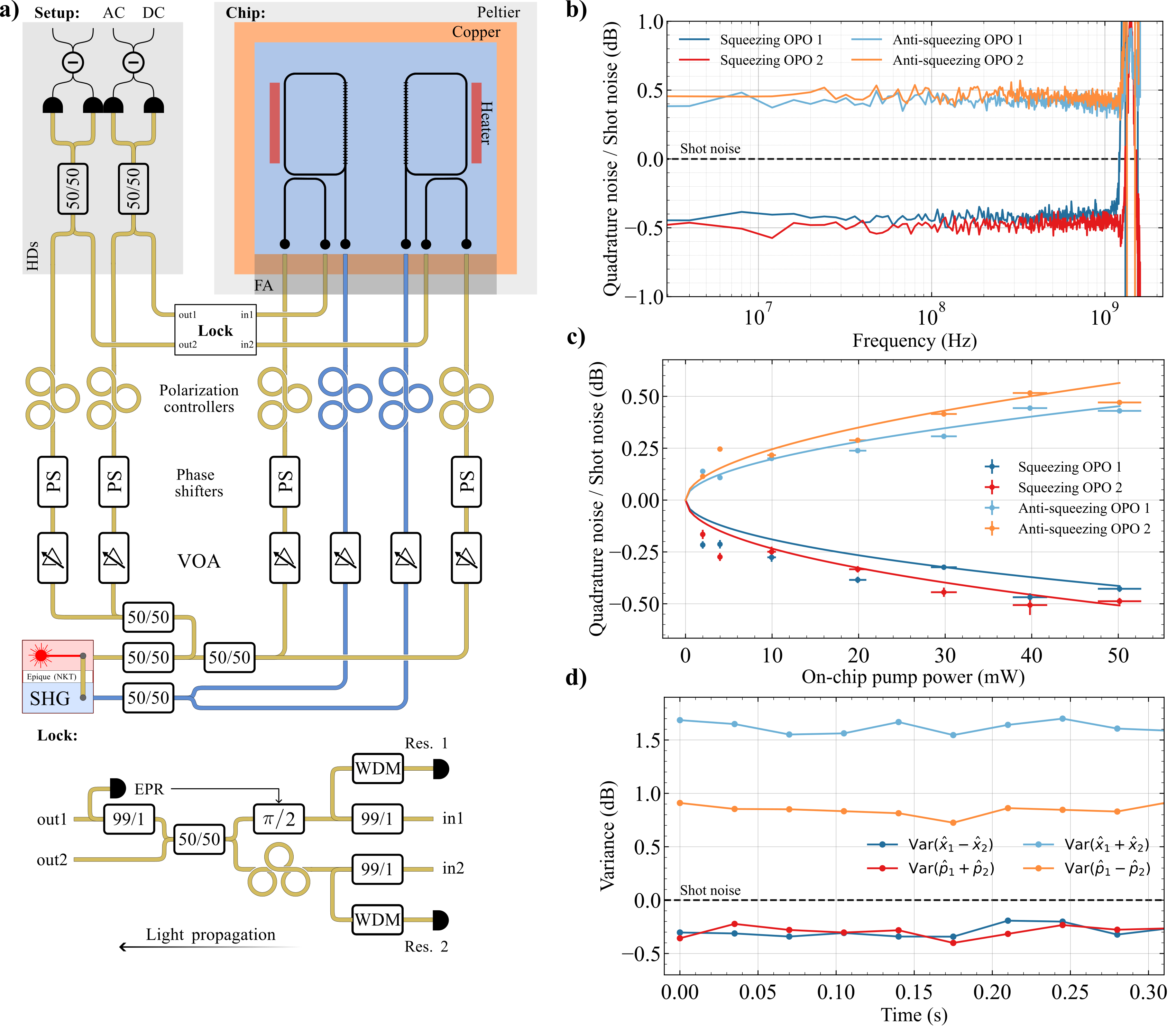}
\caption{\textbf{a)} Schematics for two-mode squeezing generation from the integrated circuit. \textbf{Setup} The fiber-based setup contains the laser source and the light distribution of both wavelength bands at \(1550\) (yellow) and \(775~\mathrm{nm}\) (blue) wavelengths. Light is manipulated using VOAs, phase shifters and polarization controllers. \textbf{Chip} Two seed and two pump sources are passed to the integrated circuit, using a fiber array for the optical and an electrical probe for the electrical signals. The global temperature is controlled with a peltier element. The output squeezed states are used for the locking scheme and passed to the HDs, where they are mixed with two LOs coming from the laser source. \textbf{Lock} The locking scheme of the setup. Both individual squeezed light sources are locked to deamplification. An additional lock controls their relative phase to create two-mode squeezing.
\textbf{b)} Spectrum obtained from the power spectral density of a 10~ms time trace of squeezing produced by the two OPOs. Dark, shot, and signal noises are sequentially acquired. After dark noise subtraction, the signals are evaluated relative to the shot noise. Squeezing and anti-squeezing show a flat spectrum up to the roll-off of the homodyne detection at 1.2~GHz. 
\textbf{c)} Investigation of the dependence of $S\pm$ on the pump power at a sideband frequency of 12~MHz. The solid lines correspond to fits of Eq.~(\ref{eq.sqz}) to obtain the power thresholds of the OPOs.
\textbf{d)} Two-mode squeezing and entanglement measurements after locking both squeezed light sources. Variances of the combinations of the quadratures $\hat{x}$ and $\hat{p}$ are shown as a function of measurement time. Continuous suppression of $\mathrm{Var}(\hat{x}_1 - \hat{x}_2)$ and $\mathrm{Var}(\hat{p}_1 + \hat{p}_2)$ below shot noise demonstrates stable two-mode squeezing. 
}
\label{fig:setup}
\end{figure*}

\subsection{Experimental setup}
The experimental setup used in this work is depicted in Fig.~\ref{fig:setup} \textbf{a)}. A two-color laser source, custom-designed and developed by NKT Photonics, delivers phase-coherent light beams at wavelengths of \(1550\) and \(775~\mathrm{nm}\) via a second-harmonic generation module. A 50/50 fiber beam splitter (FBS) splits the pump light into two paths, serving as phase-locked pumps for the two OPOs on the chip. The setup allows for independent attenuation using variable optical attenuators (VOAs) and polarization control of both input beams. The entire optical setup is built using single-mode fibers for both the pump and signal paths.

The laser also provides a tap at \(1550~\mathrm{nm}\) wavelength that is used simultanously as a seed to phase lock the resonators and as a local oscillator (LO) for homodyne detection (HD). An equal amount of light power is sent to both the seed and LO paths with additional power control achieved through electronically controlled VOAs via software. This feature facilitates the control and locking of the seeds and LOs for quantum measurements. In the seed paths, fiber-based phase shifters (PSs) are installed to allow for phase modulation, which is necessary to lock the phase of the PDC process under specific conditions. Furthermore, polarization controllers ensure that the PDC is optimized for type-0 quasi-phase matching working at the transverse electric polarization (TE). In the LO paths, the PSs and polarization controllers are employed to generate the dither signals required for phase locking of the HDs and to match the polarization between the LOs and seed signals coming from the resonators, respectively. 

The chip is placed on top of a thin copper layer that efficiently transmits heat from the Peltier element beneath it. The Peltier allows for control of the global temperature of the chip, which is used to optimize the phase matching condition at the desired wavelength. For additional control of the individual OPO resonances, the two micro-heaters are used. 

The coupling between optical fibers and the chip is achieved by using standard single-mode fibers for both wavelength bands. At the telecommunication band, G652D fibers are employed with a mode field diameter (MFD) of 10.4~µm, while HP780 Nufern (MFD = 5~µm) fibers are used at the visible band. All fibers are cleaved, assembled into a fiber array (FA), and polished. To minimize back reflections at the interface, the end facet of the FA are coated with an anti-reflection coating optimized for both bands. Handling of the different mode propagation of both wavelengths is achieved by individual design of the  polymer-based 3D printed microlenses on the chip \cite{Lohmann:25}.

Two of the fibers in the FA contain the squeezed states of light generated by the chip. These outputs are directed to two similar HDs for measurements.
All light controls for the HDs, such as polarization and phase shifting, are implemented in the LOs' paths to avoid adding losses in the squeezed light paths. The AC component of the photocurrent subtraction between the two photodiodes is recorded using a high-speed digitizer with a bandwidth of \(1.6~\mathrm{GHz}\). The DC component is used for balancing the photocurrent subtraction and also as a feedback signal for phase-locking the LOs. 
The HDs show a shot-noise clearance of around 10~dB when approximately 1.5~mW of local oscillator power is applied right before the 50/50 FBS at the detection stage. Each detector has a bandwidth of 1.2~GHz, which, combined with the large bandwidth of the digitizer, allows for broadband characterization of the squeezing spectrum.

The \textbf{Lock} part of Fig.~\ref{fig:setup} \textbf{a)} represents a schematic of the fiber configuration and locking scheme used for the generation of two-mode squeezing. This section of the setup, indicated as a white box in the \textbf{Setup} part of the figure, is connected to the two output fibers of the FA containing squeezed states. A small fraction (\(1\mathrm{\%}\)) of the seed power is tapped and passed through a wavelength division multiplexer (WDM) for filtering out any residual pump light. The infrared outputs of the WDMs are used to generate the feedback signals required for phase-locking. All monitor detectors used for locking purposes are custom-built with ultra-high-gain.

The locking sequence is as follows. First, the two OPO resonances are thermally tuned to the seed wavelength using the on-chip heaters. Second, the seed phases are locked to the deamplification points of the two OPO processes, producing phase-squeezed outputs. Third, the homodyne LO phases are independently locked to select the desired quadratures. Finally, the relative phase between the two squeezed outputs is locked to \(\pi\)/2 before the 50/50 beam splitter, which converts the two single-mode squeezed states into a two-mode squeezed state.

\subsection{Controlled generation of squeezed states}
Single-mode squeezing characterizations are performed to investigate the performance of the individual sources. 
First, pump and seed beams are sent to the resonators while scanning the phase of the PSs in the two seed paths. Both OPOs are set on resonance using the on-chip heater and the resulting amplification and deamplification patterns are observed at the output ports for different chip temperatures. The gains at each step are compared by observing the two  amplitude modulated outputs in monitor detectors connected to an oscilloscope. 
An optimal temperature of 60~\(^\circ\)C is found, which corresponds to bringing both peaks of the SHG process shown in Fig.~\ref{fig:classical results}~\textbf{d)} close to the desired wavelength. Optimal squeezing is found at 1550.12~nm wavelength, which is used for the experiment. 

After optimizing the global temperature and both on-chip heaters, quadrature noise measurements are performed in the time domain using the HDs. A possible impact of leakage of the pump light to the HD setup is ruled out, since the shot-noise level remains unchanged when the pump is injected out of resonance. A dark and shot noise measurement is acquired. For the dark noise, a time trace of both digitizer channels is acquired when all beams are blocked. Similar traces are acquired for the shot noise, where pumps and seeds are blocked, and only the LOs are incident on the photodiodes of the HDs after balancing the DC outputs. Immediately afterwards, the traces for squeezing and anti-squeezing are recorded. The electronic dark-noise spectrum is subtracted from all measured noise spectra, after which the squeezing and anti-squeezing spectra are normalized to the corresponding shot-noise spectrum. All squeezing values reported below are therefore electronic-noise-subtracted but not corrected for optical loss.
A typical noise spectrum showing squeezing and anti-squeezing from both resonators is shown in Fig.~\ref{fig:setup}~\textbf{b)}. For this measurement, a pump power of approximately 120~mW right before the FA input is used, corresponding to approximately 50~mW inside the waveguide. Higher pump powers were not investigated to avoid potential damage to the polymer couplers. The phases of the seeds are dithered and locked at the deamplification of the PDC processes to produce phase squeezing at the output of the OPOs~\cite{Wu1987_squeezed_OPO}. The seeds at the chip outputs are then optically mixed with the LO carrying another dithered signal at a different frequency and demodulated from the HD DC for selective quadrature-basis measurements. The mixing is performed using a fiber-splitter, ensuring nearly perfect visibility. 

Fig.~\ref{fig:setup}~\textbf{b)} shows a flat squeezing spectrum, consistent with the large resonators' bandwidths. Both OPOs are capable of creating squeezed light at the same wavelength with average squeezing levels at a 12~MHz sideband of -0.43 and -0.5~dB and antisqueezing of 0.43 dB and 0.47 dB for OPOs 1 and 2, respectively. The near symmetry between squeezing and anti-squeezing indicates operation well below threshold \(\sqrt{P_{pump}/P_{th}} \ll 1\) where the variances scale linearly with \(\sqrt{P_{pump}}\): \( S_\pm \approx 1 \pm 4 \eta_{tot} \sqrt{P_{pump}/P_{th}} \) (See Eq.~(\ref{eq.sqz})). In this limit, the total efficiency \(\eta_{tot}\) scales only the magnitude of the noise, and higher-order terms responsible for the large anti-squeezing asymmetry near threshold are negligible.

The dependence of \(S_{\pm}\) on the pump power is further investigated and recorded data is shown in Fig.~\ref{fig:setup}~\textbf{c)},  evaluated at the offset frequency of 12~MHz. The uncertainties in the quadrature noise are standard deviations obtained from 10 events of the same quadrature measurement acquired sequentially in intervals of 10~ms. Each data point corresponds to the minimum squeezing and maximum anti-squeezing obtained for each set. The solid lines correspond to fits of  Eq.~(\ref{eq.sqz}) using the calculated parameter for $\eta_{tot}$ mentioned in the last section and power thresholds of \(P_{th}^{(1)} = 23 \pm 4~\mathrm{W}\), and \(P_{th}^{(2)} = 15 \pm 3~\mathrm{W}\) are obtained for both OPOs respectively. The bandwidth contribution is neglected due to the flatness of the spectrum.
\subsection{Entanglement}

With both squeezers locked to deamplification of the seed, an additional FBS with a $\pi/2$ PS in one input arm is employed to create a two-mode squeezed state. The PS is actively locked using a 99/1 tap after the FBS. An additional polarization controller compensates polarization drifts in the FA and fiber connectors.
The resulting fluctuations of the sums and differences in both quadratures are shown in Fig.~\ref{fig:setup}~\textbf{d)}. 
Data is taken as synchronous time traces from both HDs. \(\mathrm{10^9}\) samples at a sample rate of 3.2~GS/s per channel are acquired which corresponds to a time duration of 312.5~ms over multiple quadrature bases. The \(10^9\) sample traces are divided into 10 blocks, each with \(10^8\) samples. Each block is digitally band-pass filtered at 10~MHz with a bandwidth of 1~MHz. The normalized variance is computed for each block corresponding to each of the data points observed in the figure. Averaging across blocks yields \(-0.292 \pm 0.017\) and \(-0.293 \pm 0.016\)~dB of squeezing for amplitude subtraction and phase sum, respectively.  Over the entire acquisition time, the noise of these quadrature combinations remains below shot noise, demonstrating the stability of the two-mode squeezing.

Continuous-variable entanglement~\cite{RevModPhys.77.513} is verified in our experiment using the Duan--Simon criterion of inseparability \cite{duan,simon}. The criterion is based on the two-mode variance of EPR-type operators. For operators \(\hat{x}_i\) and \(\hat{p}_i\) that obey the standard position-momentum commutation relations, entanglement is confirmed if variances of combinations violate the following inequality:

\begin{equation}
    \Delta^2 \left( \frac{\hat{x}_1 - \hat{x}_2}{\sqrt{2}} \right) +
    \Delta^2 \left( \frac{\hat{p}_1 + \hat{p}_2}{\sqrt{2}} \right) \geq 2.
\label{eq.entanglement}
\end{equation}

\noindent Applying the formula to the measurements in Fig.~\ref{fig:setup}~\textbf{d)}, a value of \(1.870 \pm 0.005\) is obtained, where the uncertainty corresponds to the statistical error obtained by propagating the standard deviations of the measured noise levels over time within a single acquisition through the inseparability criterion. This value violates the classical bound of 2, demonstrating EPR-type entanglement. 

\section{Conclusion}
In this work, a route towards upscaling on-chip generation of continuous-variable quantum states using thin-film lithium niobate has been demonstrated. Two independently controllable OPOs integrated on a single chip were used to generate indistinguishable squeezed vacuum states at telecom wavelengths. Individual characterization of the OPOs reveals stable squeezing levels of up to approximately 0.5 dB, while interference of both sources enabled the observation of two-mode squeezing and EPR-type entanglement, with measured correlations below the shot-noise limit.

The main limitation on the squeezing level is the relatively high OPO threshold power, \(P_{th}^{(1)} = 23 \pm 4~\mathrm{W}\) and \(P_{th}^{(2)} = 15 \pm 3~\mathrm{W}\). These high thresholds are attributed to excess propagation loss introduced during fabrication and to the deliberately strong out-coupling used to maintain a high escape efficiency. Higher squeezing levels should be achievable by reducing waveguide propagation loss, improving the nonlinear conversion efficiency of the poled section, optimizing the escape efficiency, and increasing the pump-power handling of the fiber-to-chip couplers.

Although the observed squeezing levels are modest, the key result of this experiment was the simultaneous operation of two independently tunable and mutually compatible squeezed-light sources on the same TFLN chip. The near-identical squeezing spectra and pump-power dependences indicate that the two OPOs can be operated as reproducible building blocks for larger Gaussian photonic circuits. These results represent an important step toward fully integrated continuous-variable quantum photonic circuits, combining robust and reproducible nonlinear state generation, low-loss fiber-to-chip coupling and compatibility with a larger quantum computing architectures.

Looking forward, the presented approach can be extended to larger arrays of synchronized squeezed light sources, enabling the generation of more complex circuitry required for MBQC. Next steps include deeper integration with tunable beamsplitters and electro-optical modulators on the same chip. Together with ongoing advances in integrated detectors and low-loss photonic components, this platform provides a promising route toward scalable, fault-tolerant quantum information processing on a single chip.

\begin{acknowledgments}
The authors acknowledge the support of the European Union’s Horizon Europe research and innovation program under Grant Agreement No. 101135288 (EPIQUE), No. 101080173 (CLUSTEC), and No.~101055224 (ClusterQ), as well as the Innovation Fund Denmark (PhotoQ project, no.~1063-00046A). 
The authors also acknowledge Q.ANT for the development of the homodyne detectors and NKT Photonics for providing the laser system, and especially Casper Breum for his valuable assistance and technical support.
The authors thank Jochen Stuhrmann, from Illustrato, for his assistance with the illustrations.
\end{acknowledgments}

\bibliography{bib}

\end{document}